\documentclass[aps,prd,preprint,superscriptaddress,showpacs]{revtex4}
\usepackage{epsfig}
\usepackage{dcolumn}
\usepackage{bm}
\usepackage{amsmath}
\usepackage{amsfonts}
\usepackage{amssymb}
\usepackage{graphicx}
\usepackage{latexsym}
\usepackage{multirow}
\usepackage{amssymb}
\usepackage{amsmath}
\usepackage{amsfonts}
\usepackage{mathrsfs}
\usepackage{ulem}
\usepackage{xcolor}
\usepackage{hyperref}

\begin{document}

\preprint{ACT-7-13, MIFPA-13-26}

\title{No-Scale Ripple Inflation Revisited}

\author{Tianjun Li}
\affiliation{State Key Laboratory of Theoretical Physics
and Kavli Institute for Theoretical Physics China (KITPC),
      Institute of Theoretical Physics, Chinese Academy of Sciences,
Beijing 100190, P. R. China}

\affiliation{George P. and Cynthia W. Mitchell Institute for 
Fundamental Physics and Astronomy,
Texas A\&M University, College Station, TX 77843, USA}

\affiliation{School of Physical Electronics,
University of Electronic Science and Technology of China,
Chengdu 610054, P. R. China}

\author{Zhijin Li}

\affiliation{George P. and Cynthia W. Mitchell Institute for 
Fundamental Physics and Astronomy,
Texas A\&M University, College Station, TX 77843, USA}

\author{Dimitri V. Nanopoulos}

\affiliation{George P. and Cynthia W. Mitchell Institute for 
Fundamental Physics and Astronomy,
Texas A\&M University, College Station, TX 77843, USA}

\affiliation{Astroparticle Physics Group, Houston Advanced 
Research Center (HARC), Mitchell Campus, Woodlands, TX 77381, USA}

\affiliation{Academy of Athens, Division of Natural Sciences, 
28 Panepistimiou Avenue, Athens 10679, Greece}

\begin{abstract}

We revisit the no-scale ripple inflation model, where no-scale 
supergravity is modified by an additional term for the inflaton 
field in the K\"ahler potential. This term not only breaks one
$SU(N,1)$ symmetry explicitly, but also plays an important 
role for inflation.  We generalize the superpotential in the 
no-scale ripple inflation model slightly.
There exists a discrete $Z_2$ symmetry/parity in the scalar potential
in general, which can be preserved or violated by the non-canonical
nomalized inflaton kinetic term. Thus, there are three inflation paths: 
one parity invariant path, and the left and right paths for parity 
violating scenario. We show that the inflations along the parity invariant path 
and right path are consistent with the Planck results. However, the gavitino mass
for the parity invariant path is so large that the inflation results
will be invalid if we consider the inflaton supersymmetry breaking soft mass 
term. Thus, only the inflation along the right path gives the correct
and consistent results. {\it Notably, the tensor-to-scalar ratio in such case 
can be large, with a value around 0.05, which may be probed by the future 
Planck experiment}.

\end{abstract}

\pacs{04.65.+e, 04.50.Kd, 12.60.Jv, 98.80.Cq}
\maketitle

%\newpage
% \tableofcontents
\newpage
\section{Introduction}

The first-year observations from the Planck satellite experiment on the
Cosmic Microwave Background (CMB) strongly support the six parameter
$\Lambda$CDM~\cite{Ade:2013uln}. In particular, the scalar spectral index $n_s$, 
the running of the scalar spectral index $\alpha_s \equiv {\rm d} n_s/{\rm d~ln}k$,
the tensor-to-scalar ratio $r$, and the scalar amplitude $A_s$
for the power spectrum of the curvature perturbation  are respectively 
constrained to be~\cite{Ade:2013uln} 
\begin{eqnarray}
&& n_s \simeq 0.9603 \pm 0.0073~,~~\alpha_s=-0.0134\pm 0.0090 ~,~~\nonumber \\
&& r \le 0.11~,~~A^{1/2}_s \simeq 4.6856^{+0.0566}_{-0.0628} \times 10^{-5}~.~\,
\end{eqnarray}
Also, there is no sign of 
primordial non-Gaussianity in the CMB fluctuations.
Although the Planck results are qualitatively consistent with
generic predictions of the cosmological inflationary paradigm,
many previously popular inflation models are challenged. For
example, single field inflation models with a monomial potential $\phi^n$
for $n\ge 2$ are disfavoured. Interestingly, 
the Starobinsky $R+R^2$ model~\cite{Staro, MC}
predicts a value of $n_s \sim 0.96$, which is in perfect agreement 
with the CMB data, and a value of $r \sim 0.004$ that is comfortably 
consistent with the Planck upper bound~\cite{Ade:2013uln}.

On the other hand, it is well-konwn that supersymmetry is the most 
promising construct for new physics beyond the Standard Model (SM) 
of particle physics. 
Especially, it can stabilize the scalar masses, and the superpotential is 
not renormalized. Given also that gravity is very important in the 
early Universe, it seems to us that a natural framework for
inflation model building is supergravity~\cite{SUGRA}. However,  
the supersymmetry breaking scalar masses in a generic supergravity theory
are of the same order as the gravitino mass, giving rise to 
the so-called $\eta$ problem~\cite{eta}, 
where all the scalar masses are of the order of the Hubble parameter
because of the large vacuum energy density during inflation~\cite{glv}.
Although this problem can be solved for inflationary models within
 simple supergravity~\cite{nost,hrr}, the solution relies on a seemingly
accidental cancellation in the inflaton mass~\cite{lw}.

No-scale supergravity was proposed to solve the cosmological 
flatness problem~\cite{Cremmer:1983bf}.
It satisfies the following three constraints:
(i) The vacuum energy vanishes automatically due to
a suitable chosen K\"ahler potential; (ii) At the minimum of
the scalar potential, there are flat directions that leave
the gravitino mass $M_{3/2}$
undetermined; (iii) The quantity ${\rm Str}{\cal M}^2$
is zero at this minimum. If the third condition were not applicable, 
large one-loop corrections would force the 
gravitino mass to be either identically zero or of the Planck scale.
Interestingly,  no-scale supergravity can be realized naturally
in the compactifications of weakly coupled heterotic string
theory~\cite{Witten:1985xb} and M-theory on $S^1/Z_2$~\cite{Li:1997sk}. 
Crucially, quadratic scalar mass terms are suppressed, and 
the effective scalar potential is similar to that occuring in
 global supersymmetric
models, such that the $\eta$ problem is solved naturally.
Therefore, no-scale supergravity is a very important framwork 
for inflation~\cite{Ellis:1984bf, Enqvist:1985yc}. 
Most remarkably, it has been shown recently that 
the cosmological inflation models resembling 
the Starobinsky $R + R^2$ model can be constructed naturally in no-scale 
$SU(N,1)/SU(N) \times U(1)$ supergravity 
with $N > 1$~\cite{Ellis:2013xoa, Ellis:2013nxa}, which provides 
a strong link between particle physics and cosmology.
The inflaton field may be identified with either a modulus field or 
a matter field. If a matter field plays the role of the inflaton, 
the modulus field may be stabilized as well~\cite{Ellis:2013nxa}.
For subsequent related works, see Refs.~\cite{Kallosh:2013lkr, Buchmuller:2013zfa,
Kallosh:2013hoa, Farakos:2013cqa, Ferrara:2013rsa, Briscese:2013lna}.

In this paper, we will revisit the no-scale ripple inflation model,
which was proposed by Enqvist, Nanopoulos, and Quiros (ENQ)~\cite{Enqvist:1985yc}.
For simplicity, we shall call it the ENQ model in the following. In this model, 
there exists an extra term of inflaton field in the K\"ahler potential, 
which breaks $SU(N,1)$ symmetry explicitly and plays an important role
for inflation. However, the ENQ model is not consistent with the Planck
results because its tensor-to-scalar ratio is too large for 
$n_s$ around 0.96. Thus, 
we modify the superpotential in the ENQ model slightly.
In general, there is a discrete $Z_2$ symmetry/parity in the scalar potential,
 which can be preserved or violated by the non-canonical inflaton kinetic
term. Thus, we have three inflation paths: one parity invariant
path, and the left and right paths for parity violating scenario.
We find that the inflation along both the parity invariant path and the right
path is consistent with the Planck results. However, the gavitino mass
for the parity invariant path is so large that the inflation results
will not be valid if we consider the supersymmetry breaking soft mass term for the inflaton
field. Thus, only the inflation along the right path gives the correct
and consistent results. {\it Notably, the tensor-to-scalar ratio in this case 
can be large, with a value around 0.05, which may be probed by the future 
Planck experiment}.

\section{The ENQ Model}

First, let us briefly review the ENQ inflation model~\cite{Enqvist:1985yc}. 
It is based on the maximally symmetric 
supergravity with a generalized $SU(N,1)$ manifold. The K\"ahler function is
\begin{equation}
G=-3\ln \left[z+z^*-K'(\phi,\phi^*)-\frac{1}{3}y^i {y_i}^* \right]-a K'(\phi,\phi^*)+F+F^*,
\label{kp}
\end{equation}
where $z$ is the modulus field, $\phi$ is the inflaton, and $y_i$ are matter fields. 
The term $F=F(\phi,y_i)$ is the superpotential, which, in a more familiar notation, 
is expressed as $F={\rm{ln}}~W(\phi,y_i)$. The leading term 
${\cal G} \equiv -3\ln [z+z^*-K'(\phi,\phi^*)-\frac{1}{3}y^i {y_i}^*]$ is the no-scale supergravity 
K\"ahler potential if $K'(\phi,\phi^*)$ is real, 
which satisfies the flatness condition 
${\cal G}_i^{j*}\partial^i{\cal G}\partial_{j*}{\cal G}=3$. 
This type of K\"ahler potential has important applications in cosmology, 
such as inflation~\cite{Ellis:1984bf, Ellis:2013xoa, Ellis:2013nxa}. 
The no-scale supergravity inflation 
models are favored because the scalar potential, althought not 
perfectly flat after supersymmetry breaking, 
still slopes sufficiently gently to avoid the $\eta$ 
problem~\cite{eta} and provide slow-roll inflation. 
The extra term $aK'(\phi,\phi^*)$, which can be considered as higher order correction, 
breaks the $SU(N,1)$ symmetry of the K\"ahler potential and generates a non-flat
direction like a ``ripple''. Otherwise, the potential is flat 
at tree level. We will show that, with such a term, inflation can be realized in 
a relatively simple and natural way.

The kinetic terms of the scalars are given by $K_i^{j^*}\partial_\mu \phi^i \partial^\mu \phi_j^*$, 
where $K^{j^*}_i \equiv {\partial}^2 K / \partial \phi^i \partial \phi_j^* $ is 
the K\"ahler metric. Specifically, the kinetic term of the inflaton field $\phi$ is
\begin{equation}
{\cal L}_{KE}=\left(3x^2K'_\phi K'_{\phi^*}+(3x-a)K'_{\phi \phi^*} \right)\partial_\mu \phi~\partial^\mu \phi^*~,~\,
\label{ke}
\end{equation}
where $x=(z+z^*-K'(\phi,\phi^*)-\frac{1}{3}y^i {y_i}^*)^{-1}$. The scalar potential is 
given by
\begin{equation}
V=e^G \left[\frac{\partial G}{\partial \phi^i}(K^{-1})_{j^*}^i \frac{\partial G}{\partial {\phi_{j^*}}}-3 \right] ~,
\end{equation}
in which $(K^{-1})_{j^*}^i$ is the inverse of the K\"ahler metric. 
Assuming that the D-term contribution to the potential vanishes during inflation, 
 we obtain an effective scalar potential corresponding to the K\"ahler potential 
in Eq.~(\ref{kp})
\begin{eqnarray}
V&=&\frac{x^3}{(3x-a)K'_{\phi \phi^*}} {\rm{exp}}(F+F^* -aK')|F_\phi-a K'_\phi |^2 \nonumber \\
 && + x^2{\rm{exp}}(F+F^*-aK')F_i^* F^i.
\label{sp}
\end{eqnarray}
From Eqs. (\ref{ke}) and (\ref{sp}), we can see that in the region with positive kinetic energy, 
{\it i.e.}, with $3x-a>0$, the scalar potential is always positive semi-definite as long 
as $K'_{\phi \phi^*}>0$.

The potential in Eq.~(\ref{sp}) depends on the inflaton field $\phi$, matter fields $y_i$, and 
the combined term $x$. During inflation, all these terms, except the inflaton, are fixed 
in a local minimum of the potential, which requires $\frac{\partial F}{\partial y_i}=0$ 
and $\frac{\partial V}{\partial x}=0$. The second condition leads to $x=\frac {a}{2}$, 
and the matter fields $y_i$ are set to zero at the inflation energy scale. 
One can easily check that this is a stable point, i.e., 
$\frac{\partial^2 V}{\partial^2 x}>0$. From $x=\frac {a}{2}$, we obtain
the real part of modulus $z$ which varies during inflation
\begin{eqnarray}
{\rm Re} (z) &=& \frac{1}{a}+ \frac{1}{2}K'(\phi,\phi^*)~.~\,
\end{eqnarray}
Because the inflaton potential in our paper only depends on $x$, we can choose
the proper value of ${\rm Re} (z)$ so that the inflaton potential is minimized, and
 $x$ is a constant and equal to $\frac {a}{2}$  during inflation.

The scalar potential during inflation can then be simplified as follows
\begin{equation}
V=\frac{a^2}{4K'_{\phi\phi^*}}{\rm{exp}}(F+F^*-aK')|F_\phi-aK'_\phi|^2~.
\label{infp}
\end{equation}
Obviously, this potential vanishes if $a=0$, and ripple inflation is generated by 
the non-zero $a$. Also, it has a global minimum at $F_\phi-aK'_\phi=0$, since the potential
is non-negative for $K'_{\phi\phi^*} > 0$. Inflation evolves 
towards this global minimum along the valley described by Eq.~(\ref{infp}), and the 
pre-inflation state is assumed to be close to the origin.

Similar to Ref.~\cite{Enqvist:1985yc}, we will take $K'(\phi, \phi^*)$ as the canonical
K\"ahler potential term, {\it i.e.}, $K'\equiv \phi\phi^*$. The kinetic term $L_{KE}$ is
\begin{equation}
L_{KE}=(3x^2\phi\phi^*+(3x-a))\partial_\mu\phi\partial^\mu\phi,
%\label{infp}
\end{equation}
which is non-canonical and will play an important role in the inflation.
In ENQ model, the superpotential is
chosen as follows
\begin{equation}
F={\rm{ln}}m_0+\xi\phi+\frac{1}{2}(a-\xi^2)\phi^2+\cdots, 
\label{F-term}
\end{equation}
where $m_0$ is the overall energy scale, and
the higher order terms in $F$ are ignored.

\begin{figure}
\centering
\includegraphics[width=80mm, height=70mm,angle=0]{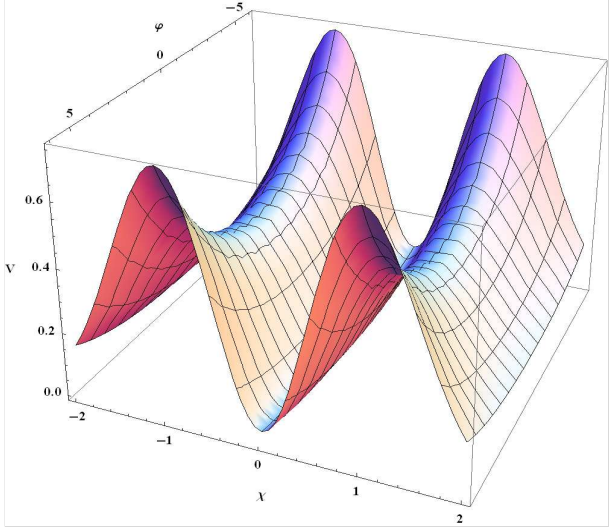}
%\hspace{-.1mm}
\caption{\label{potential-g} The inflationary potential for the parity invariant scenario with 
$\frac{b}{2a}=0.02$. During inflation the field $\varphi$ (re-scaled) runs from $4.9$ to $1.3$. 
The overall coefficient of the scalar potential is ignored as it is irrelevant to the slow-roll analysis.}
\end{figure}

In this paper, we will generalize the ENQ superpotential slightly. 
First, we consider a simple case with superpotential 
\begin{equation}
F={\rm{ln}}m_0+\frac{1}{2}(a-b)\phi^2. \label{F1}
\end{equation}
Decomposing the complex field as 
$\phi\equiv\frac{1}{\sqrt{2}}(\varphi+i\chi)$, the scalar potential becomes
\begin{equation}
V=\frac{1}{4}a^2{m_0}^2 e^{(\frac{b}{2}-a)\chi^2-\frac{b}{2}\varphi^2}
\left(2(\frac{b}{2}-a)^2\chi^2+\frac{b^2}{2}\varphi^2 \right)~.
\end{equation}
And the kinetic term is
\begin{equation}
L_{KE}=\frac{a}{2}\left(\frac{3a}{8}(\varphi^2+\chi^2 )+\frac{1}{2}\right)
\left(\partial_\mu\varphi\partial^\mu\varphi
+\partial_\mu\chi\partial^\mu\chi \right)~.
\end{equation}
Making the following transformations on $\varphi$ and $\chi$
\begin{equation}
\varphi\to\sqrt{a}\varphi~,~~~\chi\to\sqrt{a}\chi~,~\,
\label{Rescale-Eq}
\end{equation}
we obtain the scalar potential and kinetic term 
\begin{equation}
V=\frac{1}{2}a^3{m_0}^2 e^{(\frac{b}{2a}-1)\chi^2-\frac{b}{2a}\varphi^2} \left((\frac{b}{2a}-1)^2\chi^2
+(\frac{b}{2a})^2\varphi^2 \right)~,
\label{pei}
\end{equation}
\begin{equation}
L_{KE}=\frac{1}{2}\left(\frac{3}{8}(\varphi^2+\chi^2)+\frac{1}{2} \right)
\left(\partial_\mu\varphi\partial^\mu\varphi
+\partial_\mu\chi\partial^\mu\chi \right)~.~ \label{kee1}
\end{equation}
Thus, the scalar potential and kinetic term are invariant under the following
discrete $Z_2$ symmetry or parity
\begin{equation}
\varphi \leftrightarrow \chi~,~~~\frac{b}{2a}-1\leftrightarrow -\frac{b}{2a}~.~\,
\end{equation}
Therefore, we can consider either the real part $\varphi$ or imaginary part $\chi$ as the inflaton.
The above potential has a global minimum $V(\varphi,\chi)=0$ at 
$\varphi=\chi=0$, where the masses of fields $\varphi$ and $\chi$ are
\begin{equation}
\begin{split}
m_\chi^2=V_{\chi\chi}(0,0)=a^3{m_0}^2 \left(\frac{b}{2a}-1 \right) ^2~, \\
m_\varphi^2=V_{\varphi\varphi}(0,0)=a^3{m_0}^2 \left(\frac{b}{2a} \right)^2~.  ~~~~~
\end{split}
\end{equation}
Without loss of generality, we assume that the real part $\varphi$ plays the role of inflaton 
(up to a field redefinition for canonical normalization), and the imaginary part $\chi$ is stabilized 
at the minimum. The masses of the two fields should satisfy the condition $m_\varphi^2\ll m_\chi^2$, 
or equally $|\frac{b}{2a}|\ll1$. This constraint can be satisfied in the numerical results based 
on the Planck observations. An example for the scalar potential with $\frac{b}{2a}=0.02$ 
is shown in Fig.~\ref{potential-g}. It is obvious that the inflaton $\varphi$ provides 
the relatively much flatter direction while the direction along $\chi$ is very steep.

The potential with $\chi=0$ for the inflation is 
\begin{equation}
V=\alpha \varphi^2 e^{-\frac{b}{2a}\varphi^2} , \label{pe1}
\end{equation}
in which the coefficient $\alpha=\frac{1}{8}ab^2{m_0}^2 $ is not important for the slow-roll analysis. 
Note that we have $\frac{b}{2a}\ll1$ and $\varphi <\frac{2a}{b}$ during inflation,
we obtain the field $\chi$ mass along the inflation path 
\begin{equation}
\begin{split}
m_\chi^2=V_{\chi\chi}(\varphi,0)\simeq a^3 {m_0}^2 \left(\frac{b}{2a}-1 \right) ^2 e^{-\frac{b}{2a}\varphi^2}+\cdots \\
\simeq \frac{2\left(\frac{b}{2a}-1 \right) ^2 }{(\frac{b}{2a})^2\varphi^2} V(\varphi,0)\geqslant6H^2~,~~~~~~~
\end{split}
\end{equation}
where a negative mass term proportional to $(\frac{b}{2a})^2 \varphi^2$ has been dropped since it
gives small contribution to the $\chi$ mass.  Therefore, the field $\chi$ is stabilized during inflation.

\begin{figure}
\centering
\includegraphics[width=80mm, height=50mm,angle=0]{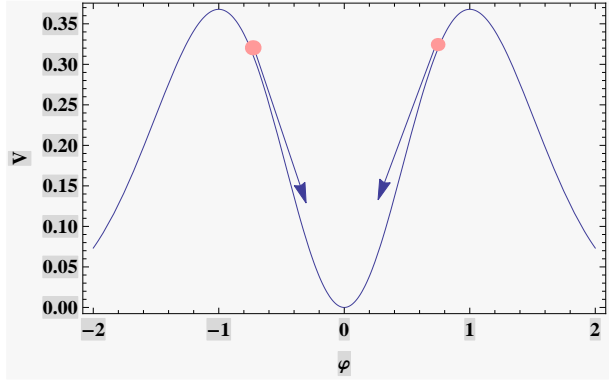}
%\hspace{-.1mm}
\caption{\label{fig1}  The potential $e^{-\varphi^2} \varphi^2$ with parity invariant scenario. The inflations along 
the left and right paths are the same.}
\end{figure}

Also, the kinetic term $L_{KE}$ becomes
\begin{equation}
L_{KE}=\frac{1}{2} \left(\frac{3}{8}\varphi^2+\frac{1}{2} \right)\partial_\mu\varphi\partial^\mu\varphi~. \label{ke1}
\end{equation}
Both the potential and kinetic term are function of $\varphi^2$, consequently, the potential after canonical 
normalization should be parity invariant, and has the global minimum at $\varphi=0$. The inflations along 
the left and right sides are exactly the same due to the $Z_2$ symmetry/parity, as shown in Fig.~\ref{fig1}.

\begin{figure}
\centering
\includegraphics[width=80mm, height=50mm,angle=0]{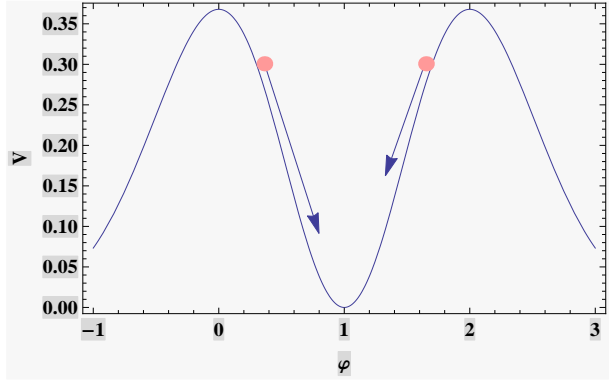}
%\hspace{-.1mm}
\caption{\label{fig2} The potential $e^{-\lambda(\varphi-\mu)^2} (\varphi-\mu)^2$ with symmetry 
along new axis $\varphi=\mu$ where $\mu$ is normalized to 1. 
However, the inflation along the left and right paths are totally different due to the canonical normalization.}
\end{figure}

Next, we consider the general superpotential $F$ with an additional linear term
\begin{equation}
F={\rm{ln}}m_0+\frac{1}{2}(a-b)\phi^2+\frac{c}{\sqrt{2}}\phi~.~ \label{F2}
\end{equation}
So the scalar potential turns into
\begin{equation}
V=\frac{1}{2}a^3{m_0}^2 e^{\frac{c^2}{2b}} e^{(\frac{b}{2a}-1)\chi^2-\frac{b}{2a}(\varphi-\frac{\sqrt{a}c}{b})^2}
\left((\frac{b}{2a}-1)^2\chi^2+(\frac{b}{2a})^2(\varphi-\frac{\sqrt{a}c}{b})^2 \right)~,
\end{equation}
in which the fields have been rescaled as in Eq.~(\ref{Rescale-Eq}).
After introducing the linear term in the superpotential, we obtain the general $Z_2 $ discrete  symmetry 
or parity for scalar potential
\begin{equation}
\varphi' \leftrightarrow \chi~,~~~\frac{b}{2a}-1\leftrightarrow -\frac{b}{2a}~,~\,
\label{G-DS-Eq}
\end{equation}
where 
\begin{equation}
\varphi' ~=~\varphi-\frac{\sqrt{a}c}{b}~.~\,
\end{equation}
 As before, with a tiny value of $|\frac{b}{2a}|$, the field $\chi$ obtains mass much larger than $\varphi$ and then
is fixed at the global minimum. The corresponding scalar potential is
\begin{equation}
V=\frac{1}{8}ab^2{m_0}^2 e^{\frac{c^2}{2b}} e^{-\frac{b}{2a}(\varphi-\frac{\sqrt{a}c}{b})^2} 
\left(\varphi-\frac{\sqrt{a}c}{b} \right)^2,
\end{equation}
or more concisely,
\begin{equation}
V=\alpha e^{-\lambda(\varphi-\mu)^2} (\varphi-\mu)^2. \label{pe2}
\end{equation}
The potential in ENQ's model is equal to the case with $\lambda=\frac{b}{2a}=\frac{1}{\mu^2}$, 
which gives the constraint $\frac{c^2}{2b}=1$ and $\alpha=\frac{1}{8}ab^2{m_0}^2 e$. 
The potential in Eq.~(\ref{pe2}) is the horizontal shift of the potential in Eq.~(\ref{pe1}),
and is still symmetric under a new axis $\varphi=\mu$, as shown in Fig.~\ref{fig2}.
Because the kinetic term is the same as the above simple case, the discrete symmetry or parity
is broken by it. So after the canonical normalization, the symmetry 
with respect to the axis $\varphi=\mu$ disappears, and then the inflations along the left and 
right paths are significantly different!

\section{Inflation Study}

\begin{figure}
\centering
\includegraphics[width=90mm, height=60mm,angle=0]{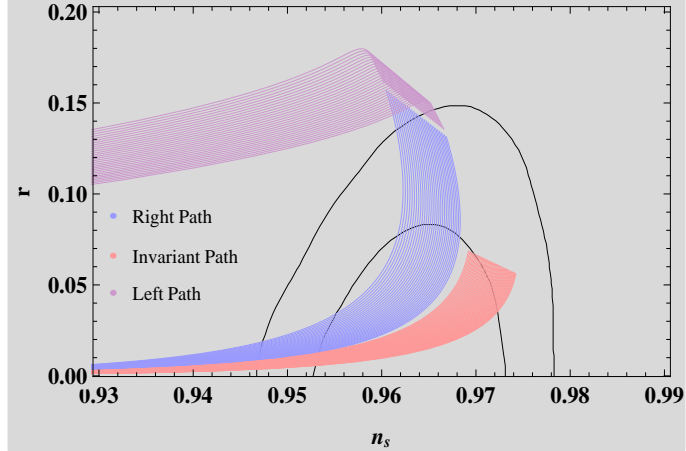}
\caption{Numerical results for the inflations along the parity invariant path and left/right asymmetrical paths. 
The parity invariant inflation is close to the result of inflation with potential $\alpha\phi e^{-\beta\phi}$, 
while the right path is similar to the case with potential $\alpha\phi^2 e^{-\beta\phi}$. The left path is 
out of $2\sigma$ 
region of the Planck observations. Inflations along the left and right paths coincide with each other at 
the approximation of the chaotic inflation model with $V=\frac{1}{2}{m_0}^2\phi^2$.} 
\label{fig3}
\end{figure}

Taking different superpotentials, we get three different inflation paths: the parity invariant path 
and the asymmetrical left and right paths. The inflation predictions along three paths for 
the canonical normalized inflaton field are given
in Fig.~\ref{fig3}. For each strip in Fig.~\ref{fig3}, the upper and lower boundaries correspond
 to the e-folding numbers $N=50$ and $N=60$, respectively. Besides the e-folding number $N$, 
the $n_s-r$ relation only depends on the parameter $\lambda=\frac{b}{2a}$. 
Thus, the strip ranges are obtained by varying  $\lambda$ with different $N\in[50, 60]$.
By the way, the parameter $\alpha$ can be determined by the observed scalar amplitude 
via $A_s= \frac{2V(\varphi_i)}{3\pi^2 r}$  in Planck unit~\cite{Ade:2013uln}, and it
is about $1.5\times 10^{-11}$~\cite{Li:2013nfa}. Interestingly,
the tensor-to-scalar ratio is larger than $0.01$,
which is above the Lyth bound~\cite{Lyth:1996im} and belongs to the large field inflation class. 
Among the three paths, inflation along the parity invariant and right paths
can fit the Planck observations. 

\subsubsection{Inflation along the Parity Invariant Path}

The potential in Eq.~(\ref{pe1}) as well as the kinetic term is invariant under
the parity, so the inflations along 
the left and right valleys are the same. The main predictions for $n_s$ and $r$ are within the $1\sigma$ region
of the Planck observations. Specifically, for $n_s=0.9603$ and e-folding number $N=55$, the tensor-to-scalar ratio $r$ 
is about $0.02$, and the re-scaled field $\varphi$ runs from $\varphi_i=4.9$ to $\varphi_f=1.3$. 
The coefficient $\frac{b}{2a}\simeq0.026$, which is much smaller than $1$, validates 
the initial assumption for the stabilization of the imaginary part $\chi$.

Along this path the inflation is similar to the inflationary results of the potential $\alpha\phi^ne^{-\beta^m\phi^m}$ 
with $m=1$ and $n=1$. Because the inflation is the typical large field inflation,  the kinetic term approximately is
\begin{equation}
\begin{split}
L_{KE}=\frac{1}{2}(\frac{3}{8}\varphi^2+\frac{1}{2})\partial_\mu\varphi\partial^\mu\varphi   \\ \label{ke2}
\simeq\frac{3}{16}\varphi^2 \partial_\mu\varphi\partial^\mu\varphi~.~~~~~~~~~
\end{split}
\end{equation}
The canonical normalization is close to $\varphi\to\sqrt{\frac{3}{32}}\varphi^2\equiv\psi$, so
the corresponding scalar potential with canonical normalized inflaton field is 
\begin{equation}
V=\alpha e^{-\frac{b}{2a}\varphi^2} \varphi^2=\alpha \psi e^{-\beta \psi}.
\end{equation}
Therefore, the parity invariant path is close to the inflation with potential 
$\alpha\phi e^{-\beta\phi}$~\cite{Li:2013nfa} in the large field approximation.

\subsubsection{Inflation along the Left and Right Paths}

The inflations along the left and right asymmetrical paths are interesting  
due to the non-canonical kinetic term. Without canonical normalization of the inflaton field, 
the potential is still symmetric between the two sides of the global minimum. 
However, this symmetry is broken by the non-canonical inflaton kinetic term. 
As a result, the inflation processes along the two paths are 
significantly different. As shown in Fig.~\ref{fig3}, the left path gives too large $r$ which is out of 
the $2\sigma$ region of the Planck results. However, for the right path, it gives totally different results, 
which highly agree with the Planck observations. Here, for the linear term coefficient $c$, we adopt 
the value used in ENQ's paper so that $\lambda=\frac{1}{\mu^2}$, or $\frac{c^2}{2b}=1$.
The right path gives the results with very large range of $r$, from $0.02$ to $0.08$ within $1\sigma$ region. 
Specifically, for $n_s=0.9603$ with e-folding number $N=55$, the tensor-to-scalar ratio $r$ is about $0.05$, 
the re-scaled field $\varphi$ runs from $\varphi_i=7.7$ to $\varphi_f=5.1$. The coefficient $\frac{b}{2a}\simeq0.047$ 
is also much smaller than $1$, therefore, the condition to stabilize field $\chi$ is satisfied.

There is an interesting fact for the left and right paths: the two strips for the left and right paths are going to 
coincide, and even perfectly fit with each other at the end of the two strips. Actually, this is the prediction 
of the chaotic inflation model with $V=\frac{1}{2}{m_0}^2\phi^2$! Let us explain it briefly.
The inflation exit condition for the parity violating scenario is
\begin{equation}
\epsilon=\epsilon'\frac{1}{\frac{3}{8}\varphi_f^2+\frac{1}{2}}=1, \label{exit}
\end{equation}
in which the $\epsilon'$ is 
\begin{equation}
\epsilon'=\frac{1}{2}(\frac{V_\varphi}{V})^2~,~
\end{equation}
and has a pole at the global minimum $\varphi_0=\mu$. For large field $\varphi_f>1$, the exit condition 
in Eq.~(\ref{exit}) 
requires a large $\epsilon'$, which can be realized only in the region close to the pole $\varphi_0=\mu$.
 In this region, the potential can be expanded in terms of the canonical normalized field $\psi$ as
\begin{equation}
V(\psi)=V(\mu')+V'(\mu')(\psi-\mu')+\frac{1}{2}V''(\mu')(\psi-\mu')^2+\cdots,
\end{equation}
where $\mu'$ is the value of $\psi$ at the global minimum.
Here, $V(\mu')=V'(\mu')=0$, $V''(\mu')={m_0}^2$ and $\mu'$ is large. Therefore, in this approximation, 
no matter we approach the global minimum from left or right, the inflation is the same as the 
 chaotic inflation with $V=\frac{1}{2}{m_0}^2\phi^2$.

\begin{figure}
\centering
\includegraphics[width=90mm, height=60mm,angle=0]{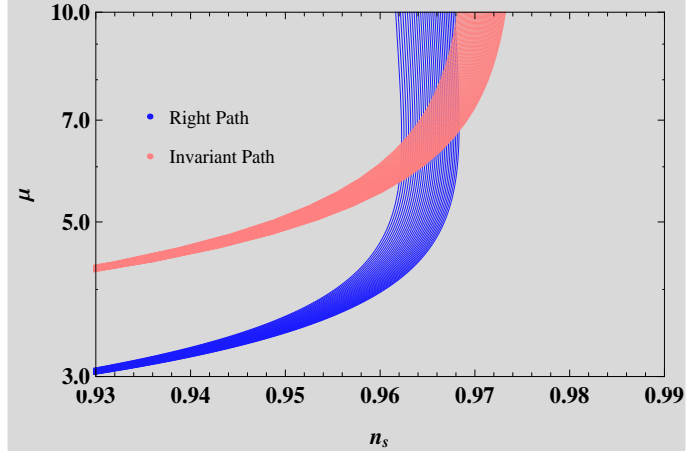}
\caption{Value of the shift parameter $\mu$ versus  the spectral index $n_s$. For the $n_s\sim0.9603$, 
the shift parameter $\mu$ varies from 5-10 (6-8) for the right (invariant) path, 
which corresponds to the parameter $\frac{b}{2a}\sim O(10^{-2})$. } \label{fig4}
\end{figure}

\subsubsection{Stabilization Condition}

To get the single field inflation, the extra field $\chi$ has been stabilized by the condition 
that its mass is much larger than the inflaton $\varphi$. The condition is $\frac{b}{2a}\ll1$ 
so that the mass of $\varphi$ is several orders smaller than $\chi$. Fig.~\ref{fig4} shows 
the relation between the spectra index $n_s$ and the shift parameter $\mu$, 
from which the value of the parameter $\frac{b}{2a}=\frac{1}{\mu^2}$ can be evaluated. 
It is shown that close to the central value of $n_s\sim0.9603$, the parameter $\frac{b}{2a}$ 
is about ${\rm O}(10^{-2})$, which leads to the mass ratio $\frac{m_\varphi^2}{m_\chi^2}\sim10^{-4}$.
Therefore, the ENQ model can stabilize the extra field $\chi$ and fit the Planck observations simultaneously. 
While for the right path with much smaller $n_s<0.93$, $\mu\leq3$ and then $\chi$ can not be stabilized.

\section{Comments on Gravitino Mass}

Because the vacuum corresponds to the large field $\varphi>1$ in the Planck unit,
the gravitino mass will be unacceptable large if there does not exist significant cancellations.
Thus, the gravitino mass needs to be considered carefully.

The gravitino mass $M_{\frac{3}{2}}$ is
\begin{equation}
M_{\frac{3}{2}}={\rm exp} \left(\frac{<G>}{2} \right)=x^{\frac{3}{2}} \times
{\rm exp} \left(\frac{<-a K'(\phi,\phi^*)+F+F^*>}{2} \right)~,~ \label{gravitino}
\end{equation}
in which the terms $<...>$ are evaluated at the vacuum (global minimum). For the parity invariant 
and violating scenarios, 
the vacuum expectation values of the complex inflaton field $\phi$ are respectively $0$ and $\frac{c}{b}$ with 
the constraint $x>\frac{a}{3}$. Thus, we have $<-a K'(\phi,\phi^*)+F+F^*>=2~{\rm ln}~m_0$ for 
the parity invariant scenario
and $<-a K'(\phi,\phi^*)+F+F^*>=1+2~{\rm ln}~m_0$ for the parity violating scenario. The inflation is
determined by the condition $x=\frac{a}{2}$. At the vacuum that corresponds to the end point of the inflation,
the gravitino mass is $M_{\frac{3}{2}}=m_0(\frac{a}{2})^{\frac{3}{2}}$ for the parity invariant scenario
and $M_{\frac{3}{2}}=m_0 e^{\frac{1}{2}} (\frac{a}{2})^{\frac{3}{2}}$ for the parity violating scenario.

The scalar amplitude $A_s$ for the power spectrum of the curvature perturbation from
the Planck data~\cite{Ade:2013uln} is
\begin{equation}
A_s=\frac{V}{24\pi^2\epsilon}\simeq2.196 \times 10^{-9}~, \label{As}
\end{equation}
where the inflaton is taken as $\varphi=\varphi_i$. For the parity invariant scenario, 
using $\frac{b}{2a}=\frac{1}{\mu^2}$, the potential is
\begin{equation}
\begin{split}
V=\frac{1}{8}ab^2{m_0}^2 e^{-\frac{b}{2a}\varphi^2} \varphi^2={m_0}^2(\frac{a}{2})^3
\frac{4}{\mu^2}(\frac{\varphi}{\mu})^2e^{-(\frac{\varphi}{\mu})^2} ~\\
=M_{\frac{3}{2}}^2\frac{4}{\mu^2}(\frac{\varphi}{\mu})^2e^{-(\frac{\varphi}{\mu})^2}~. ~~~~~~~~~~~~~~~~~~~~~~~~~~~~~~~~~~~
\end{split}
\end{equation}
Similarly, the potential for the parity violating scenario is
\begin{equation}
\begin{split}
V=\frac{1}{8}eab^2{m_0}^2 e^{-\frac{b}{2a}\varphi^2} \varphi^2=e{m_0}^2(\frac{a}{2})^3\frac{4}{\mu^2}(\frac{\varphi}{\mu}-1)^2e^{-(\frac{\varphi}{\mu}-1)^2}  \\
=M_{\frac{3}{2}}^2\frac{4}{\mu^2}(\frac{\varphi}{\mu}-1)^2e^{-(\frac{\varphi}{\mu}-1)^2}. ~~~~~~~~~~~~~~~~~~~~~~~~~~~~~~~~~~~~~
\end{split}
\end{equation}

\begin{figure}
\centering
\includegraphics[width=90mm, height=60mm,angle=0]{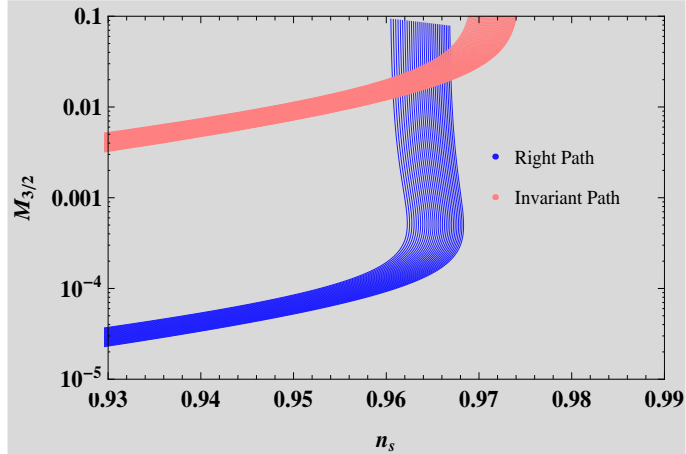}
\caption{Gravitino mass versus the spectral index $n_s$. The gravitino mass is 
significantly reduced in the right path.} \label{fig5}
\end{figure}

Applying the above equations into Eq.~(\ref{As}), the gravitino masses can be evaluated. The results are 
shown in Fig.~\ref{fig5}. For the parity invariant scenario, the gravitino mass evaluated from the inflation results 
is about $O(10^{-2})$ in the Planck unit, which is too large for inflation. In particular, the inflation results
will be invalid if we consider the supersymmetry breaking soft mass term for the inflaton field. 
Interesting, the gravitino mass can be as small as $O(10^{-4})$ in the parity violating scenario.
Thus, the inflation along the right path is indeed consistent.

%%%%%%%%%%%%%%%%%%%%%%%%%%%%%%%%%%%%%%%%%%%%%%%%%%%%%%%%%%%%%%%%%%%%%%%%%%%%%%%%%%%%%%%%%%%%%%%%

%%%%%%%%%%%%%%%%%%%%%%%%%%%%%%%%%%%%%%%%%%%%%%%%%%%%%%%%%%%%%%%%%%%%%%%%%%%%%%%%%%%%%%%%%%%%%%%%

%%%%%%%%%%%%%%%%%%%%%%%%%%%%%%%%%%%%%%%%%%%%%%%%%%%%%%%%%%%%%%%%%%%%%%%%%%%%%%%%%%%%%%%%%%%%%%%%

%%%%%%%%%%%%%%%%%%%%%%%%%%%%%%%%%%%%%%%%%%%%%%%%%%%%%%%%%%%%%%%%%%%%%%%%%%%%%%%%%%%%%%%%%%%%%%%%

%%%%%%%%%%%%%%%%%%%%%%%%%%%%%%%%%%%%%%%%%%%%%%%

%%%%%%%%%%%%%%%%%%%%%%%%%%%%%%%%%%%%%%%%%%%%%%%

\section{Conclusion}

 We have revisted the ENQ model by modifying its superpotential. We found that
there is a discrete $Z_2$ symmetry/parity in the scalar potential
in general, which can be preserved or violated by the non-canonical
nomalized inflaton kinetic term.  We showed that the inflations along the parity invariant path 
and right path are consistent with the Planck results. However, the gavitino mass
for the parity invariant path is so large that the inflation results
will not be valid if we consider the supersymmetry breaking soft mass 
term for the inflaton field. Thus, only the inflation along the right path gives the correct
and consistent results. {\it Notably, the tensor-to-scalar ratio in this case 
can be large, with a value around 0.05, which may be probed by the future 
Planck experiment}.

%%%%%%%%%%%%%
%%%%%%%%%%%%%%%%%%%%%%%%%%%%%%%%%%%%%%%%%

%%%%%%%%%%%%%%%%%%%%%%%%%%%%%%%%%%%%%%%%%%%%

%%%%%%%%%%%%%%%%%%%%%%%%%%%%%%%%%%%%%%%%%%%%
\begin{acknowledgments}

This research was supported in part
by the Natural Science Foundation of China
under grant numbers 10821504, 11075194, 11135003, and 11275246, 
by the National Basic Research Program of China (973 Program) 
under grant number 2010CB833000,
and by the DOE grant DE-FG03-95-Er-40917.

\end{acknowledgments}

\end{document}